\documentclass[11pt]{article}
\usepackage {amsfonts}
\usepackage {graphicx}
\usepackage {epsfig}
\usepackage{amssymb}
\usepackage{amsmath}
\usepackage {longtable}
\usepackage[english]{babel}
\begin{document}

\title{X-RAY SOURCE BASED ON THE PARAMETRIC X-RAYS}

\author{Alexander Lobko* and Olga Lugovskaya \\
Institute for Nuclear Problems, Belarus State University
\\
11 Bobruiskaya Str., Minsk 220050, Belarus \\ {\textit{*
Corresponding author: lobko@inp.minsk.by}}}

\maketitle

\begin{center}
\begin{abstract}

Prospects of parametric x-rays (PXR) application for the
development of a tuneable quasi-monochromatic x-ray source for
medical imaging are discussed. Analysis of basic requirements for
electron accelerator shows that it must be relatively low-energy
and high-current linac. In comparison with known
ultra-relativistic cases, at low energies PXR properties will be
modified to a great extent by multiple scattering of the
electrons. PXR intensity dependence on target thickness and beam
energy are calculated taking multiple scattering into account. It
is concluded that PXR source based on real medical accelerators is
feasible and can provide x-ray flux needful for obtaining high
quality medical images.
\end{abstract}
\end{center}



\section{Introduction}

\quad Parametric x-rays (PXR) produced by a relativistic charged
particle uniformly moving through a single crystal were declared
as a very attractive radiation mechanism for x-ray source since
its theoretical prediction in 1972 \cite{1} - \cite{3} and first
experimental observation in 1985 \cite{4}. Actually, it has a
number of indisputable features, namely, high monochromaticity,
energy tuneability, polarization, directivity, and possibility to
be emitted to large angles relative to beam direction, i.e.
provide monochromatic x-rays virtually free of background. PXR
properties were studied at a number of accelerator facilities
worldwide but a PXR-based x-ray source developments are still in
progress.

Evidently, this is connected with the fact that PXR research is performed at
existing linear and annular accelerators and limited by their technical
specifications. So, we should recognize how to design a pxr-based source
that will meet the consumer's requirements for its use in industry and
medicine.

Industry and science, as consumers, essentially yield to medicine,
which represents the huge market for devices and methods for x-ray
diagnostics. Statistically, x-ray devices count for about 60\% of
medical diagnostics devices \cite{5}. Despite the development of
alternative diagnostics methods, e.g. ultrasound and magnetic
resonance imaging, the number of studies in the field of x-ray
diagnostics remains high. First of all, medicine demands decreased
irradiation doses taken together with increased resolution and
contrast. These requirements can be met if the customer will be
offered a monochromatic tuneable source, providing x-radiation
with exactly the energy required for examination of a specific
organ. For example, in mammography the maximum spectral density
lies in the range of 17-20 keV; in radiography of the chest,
extremities and head, the range of 40-50 keV is optimal; while the
50-70 keV range is used for abdomen and pelvis radiography.
Precise selection of monochromatic radiation energy will allow a
reduction of a patient dose of tens times, even in comparison with
digital radiography and computer tomography, with improvement of
x-ray image quality. However, in comparison with scientific and
industrial applications where signal accumulation is possible, in
the case of medicine the source should emit number of quanta
sufficient to provide necessary contrast of an image at the
specific value of the signal-to-noise ratio. Estimations
\cite{6,7} give minimal value of x-ray quanta necessary for
high-quality image equal to $\sim $10$^{7}$ photons/mm$^{2}$.
PXR quantum yield in a thin crystal target even at high enough
(e.g. 900 MeV) beam energy is not very large and amounts only
$\sim $10$^{-5}$-10$^{-6}$ photons/e$^{-}\cdot $sr \cite{8}.
Assuming the target to be imaged is 30 cm diameter at one meter
distance from a source, it is necessary to apply $\sim $ 0.1 A
current to get a required x-ray flux. Such currents can be
achieved only in low energy accelerators. Though theoretically PXR
may be emitted at any energy of charged particles \cite{9}, there
is a factor limiting the minimal value of beam energy for source
under discussion.
Assuming x-ray energy necessary for subtractive coronary
angiography $\varpi _{B} $=33 keV then the angle of radiation
$\theta _{B} $ in a diamond target will come to $\sim 4^{\circ}$
for (111) reflex and $\sim 7^{\circ}$ for (220) reflex that it is
enough for arrangement of a patient at 1-2 meter distance from the
target.
Now let us estimate energy ${E}_{p}$ of an electron beam providing
angular width of reflex \cite{10} $\vartheta _{ph}^{2} = \left|
{\chi _{0}^{'}} \right| + \gamma ^{ - 2} + \overline {\theta
_{s}^{2}}  $ ($\left| {\chi _{0}^{'}} \right|$ is the Fourier
component of a crystal dielectric constant, $\gamma=E_{p}/mc^{2}$,
$\overline {\theta _{s}^{2}} $ is the mean-square angle of
multiple scattering) equal $\sim 5^{\circ}$, i.e. energy at which
x-ray reflex and forward background are still separated from each
other and the radiation in which we are interested keeps all the
properties necessary for a medical source. The dominant component
here is the mean-square angle of multiple scattering $\overline
{\theta _{s}^{2}}  $ and lowest estimation of beam energy gives a
value $\sim $17 MeV. To provide more real evaluations for PXR
source based on low-energy electron beam, we should look into
multiple scattering (MS) of electrons more precisely, because MS
appeared to be the most significant factor affecting PXR intensity
and angular distribution at relatively low energies and/or thick
targets \cite{8}. In low energy cases we shouldn't use kinematics
formulae \cite{10}, where MS is considered phenomenologically. Let
us describe the multiple scattering influence on x-ray reflex
characteristics  in more detail following \cite{11}.

\section{Calculation of x-radiation properties in the multiple
scattering presence}

\quad To get an expression for spectral-angular density of
radiation with MS account, it is necessary to average the squared
modulus of all possible particle trajectories in a crystal in the
following expression \cite{12,13}

\begin{equation}
\frac{{d^{2}N_{s}} }{{d\varpi d\Omega} } =
\frac{{e^{2}Q^{2}\varpi} }{{4\pi ^{2}\hbar c^{3}}} \left|
\int\limits_{ - \infty} ^{ + \infty}  {\vec {v}\vec {E}_{\vec
{k}}^{\left( { -}  \right)s}}  \left( {\vec {r}\left( {t}
\right),\varpi } \right) exp\left( { - i\varpi t} \right)dt
\right|^{2} \label{1}
\end{equation}

The obtained expression will describe all mechanisms of the
radiation: PXR, \textit{Bremsstrahlung}, and transition radiation
in a single manner. The velocity vector is presented in the form:
$\vec {v} = \vec {v}_{0} cos\theta + v_{0} \vec {\theta} $ (where
$\vec {v}_{0} $ determines the direction of the initial velocity
of a charged particle, $\theta \equiv \left| {\vec {\theta} }
\right|$ is the the angle of multiple scattering of a particle
($\theta < < 1$), $\vec {\theta}$ is the two-dimensional vector,
$\vec {\theta } \bot \vec {v}_{0} $ and $\vec {\theta}  = 0$ at $t
\le 0$).

Averaging is followed \cite{13}, where the procedure was performed
with the help of dispersion function, satisfying the Fokker-Planck
kinetic equation in the amorphous medium. Averaging of the right
part of the equation (\ref{1}) was performed and the integral
expressions for spectral-angular distributions of the radiation,
taking into account multiple scattering for the lateral and
forward PXR maxima were derived.

Selected functions of the particle distributions on coordinates
and angles are usually applied in calculations of
\textit{Bremsstrahlung} intensity of ultra-relativistic electrons
in amorphous media. It is applicable if direction of vector of a
charged particle initial velocity does not coincide with
directions of the basic crystallographic axes or planes. Also, one
can ignore the influence of crystal structure on charged particles
trajectories and apply dispersion functions of amorphous medium
when value of initial angular divergence of a particle beam,
falling on the crystal along the direction of main
crystallographic axes or planes, satisfies the inequality $\left|
{\Delta \vec {v}_{0}}  \right|/c > \sqrt {\left( {2U_{0} /E_{p}}
\right)} $, where $U_{0}$ is the potential of atomic axis or
plane, $E_{p}$  is the particle energy. In this case during the
passage of charged particles through the crystal, orientational
effects will be expressed less strongly than in the case of a
well-collimated beam, and the averaging of (\ref{1}) over simple
dispersion functions ensures a satisfactory approach.

It is necessary to note, that in the case, when the value $\left|
{\Delta \vec {v}_{0} /c} \right| < \left( {\gamma ^{ - 2} +
\frac{{\omega _{L}^{2} }}{{\omega _{B}^{2}} }} \right)^{ - 1/2}$
($\omega _{B}^{\left( {n} \right)} = \frac{{\pi cn}}{{dsin\theta
_{B}} }$  is the Bragg frequency, $n=1,2,...$, $\theta _{B}$ is
the angle between the particle velocity $\vec {v}_{0} $ and planes
corresponding to the vector $\vec {\tau} $,  $\omega _{L}$ is the
Langmuir frequency of a crystal), it is not required to make
additional averaging in (\ref{1})  on the initial divergence of
the vector direction $\vec {v}_{0} $.

There are two factors limiting the longitudinal size of
quasi-Cherenkov radiation formation area: MS of charged particles
on atoms of substance and x-ray absorption in the medium. As in
case of radiation generation in an amorphous medium, the PXR
characteristics essentially depend on the relation between crystal
thickness along the direction of the charged particle movement
$L_{0} $\textbf{} and \textit{Bremsstrahlung} coherent length
$L_{Br} $. The theory of PXR with the phenomenological account of
MS influence on particle energy range exceeding threshold energy
$E_{tr} = mc^{2}\left|{\chi} '_{0}\right|^{ - 1/2}$ is in good
agreement with experimental results only at crystal thickness
$L_{0} < < L_{Br} $. In this case MS influence appears only as a
small addition in the intensity of generated x-ray caused by
\textit{Bremsstrahlung}. Also MS influence on PXR phase is taken
into account and interference between PXR and
\textit{Bremsstrahlung} is neglected. In the inverse case MS
essentially changes the parameters of the PXR itself.

\section{PXR characteristics dependence on crystal target
thickness}

\quad  To demonstrate MS effect on PXR characteristics, numerical
calculations were performed for following conditions: target Si
(220) in Laue geometry, $2\theta _{B} = 19^{ \circ} $, electron
beam energy 900 MeV.

\begin{figure}
\epsfxsize = 8 cm \centerline{\epsfbox{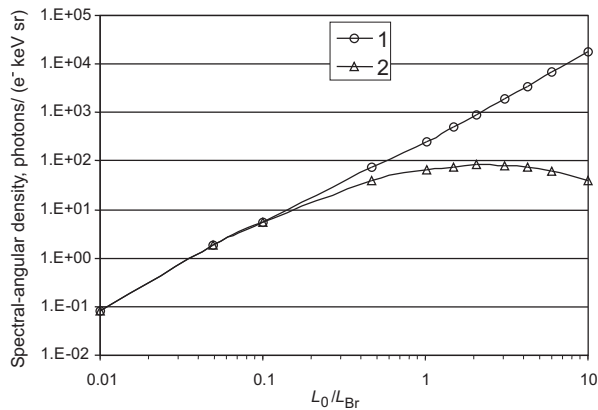}} \caption{Dependence
of PXR spectral-angular density on crystal thickness, counted
along a direction of charged particle movement in units of\textbf{
}$L_{Br} $: 1 - without MS account, 2 - with MS account}
\label{fig1}
\end{figure}

\begin{figure}
\epsfxsize = 8 cm \centerline{\epsfbox{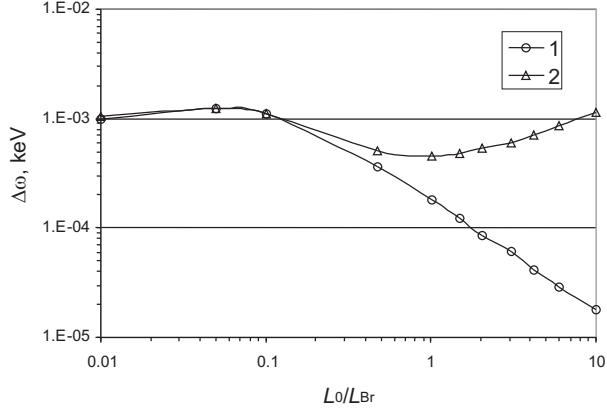}} \caption{Dependence
of PXR frequency half-width on crystal thickness, counted along a
direction of charged particle movement in units of $L_{Br} $: 1 -
without MS account, 2 - with MS account} \label{fig2}
\end{figure}

PXR spectral-angular density as a function of crystal target
thickness is shown in Fig.\ref{fig1}.
%
%
It is evident, that already at $L_{0} \sim L_{Br} $ presence of MS
results in PXR intensity distinctly less than intensity determined
without MS account. During further increase of the crystal
thickness, spectral-angular density achieves saturation at
thickness of
$\sim (2 \div 3) L_{Br}$
and then goes down.
It is seen from this figure that PXR spectral-angular density
without MS (curve 1) continues to grow with crystal thickness
increase even up to $\sim 10L_{Br}$. The tendency shows that
absorption still does not plays an essential role though $L_{0} $
becomes of the order of $L_{abs} $. Simultaneously with the
variation of PXR spectral-angular density with crystal thickness
increase, change of its spectral width appears (see
Fig.\ref{fig2}).
%
%
%

As long as the thickness of the crystal target does not exceed
$0.5L_{Br} $, spectral-angular density and spectral width of PXR
reflex determined with MS account practically coincide with the
values received without MS consideration. At the same time,
spectral width of PXR peak decreases with $L_{0} $ increase. The
width of PXR maximum is inversely proportional to radiation
coherent length. As the target thickness increase (on thickness of
$\sim L_{Br} $ and larger) PXR coherent length begins to be
limited by $L_{Br} $.
So, PXR peak width without MS continues to decrease with $L_{0}
$\textbf{} growth, but MS account results in stabilization of peak
width at $L_{0} \sim L_{Br} $ (Fig.\ref{fig2}).

Thus, calculation results, demonstrated in the Figures \ref{fig1}
and \ref{fig2}, show that MS of a charged particle results not
only in radiation phase shift, but also in essential decrease of
PXR spectral-angular density at angles of $\vartheta \sim
\vartheta _{ph} $ as well as in change of reflex width, even in
case of $L_{0} \sim L_{Br} $.

\begin{figure}
\epsfxsize = 8 cm \centerline{\epsfbox{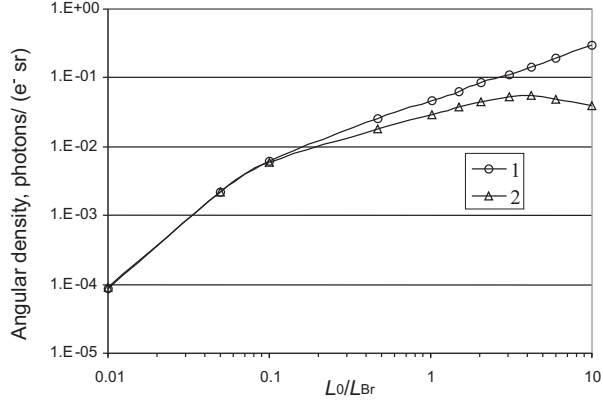}} \caption{Dependence
of PXR angular density maximum on crystal thickness, counted along
a direction of charged particle movement in units of $L_{Br} $
(integration on frequency on half-width of PXR spectral-angular
density distribution): 1 - without MS account, 2 - with MS
account} \label{fig3}
\end{figure}

As a result of joint processes of PXR spectral narrowing and
spectral-angular density decrease, PXR integral characteristics do
not vary essentially.
As an example, in Fig.\ref{fig3} there are given dependence of PXR
angular distributions (at $\vartheta = 1.66 \cdot 10^{ - 3}$rad)
on the crystal thickness without MS (1)  and in the MS presence
(2). One can see that for the target thickness $L_{0} = 10L_{Br} $
the maximal value of spectral-angular distribution (at $\vartheta
= 1.66 \cdot 10^{ - 3}$ rad) calculated without MS account exceeds
the the maximal value obtained with MS account more than two
orders of magnitude (compare with Fig.\ref{fig1}), while
amplitudes of angular distributions (for the same angle) differ
less than by one order (Fig.\ref{fig3}). MS also results in PXR
angular distribution spreading, therefore if the angular aperture
of a detector is big enough ($\theta _{D} > > \vartheta _{ph} $),
the difference in results for integrated number of the quanta
calculated with the MS account and without taking it into account
will be even less, than for angular distributions. Thus, the most
significant difference of the results obtained with and without MS
account is observed for differential distributions.
%
%
%

\section{PXR intensity dependence on charged particle energy}

\quad As well as for Cherenkov radiation in homogeneous media, PXR
spectral-angular density has threshold behavior.
 However, due
to essential dispersion of a crystal refraction index, dependence
of integral PXR intensity on particle energy is relatively smooth.
It is differs in $E_{p} < < E_{tr} $ and $E_{p}
> > E_{tr} $ energy regions, where threshold energy $E_{tr} =
mc^{2}\gamma _{tr} $, $\gamma _{tr} = \sqrt {\left| {{\chi} '_{0}}
\right|} \approx \frac{{\omega _{B}} }{{\omega _{L} }}$. So, PXR
quantum yield at particle energy $E_{p} < < E_{tr} $\textbf{ }is
$N^{PXR}\sim \left( {\frac{{E_{p}} }{{E_{tr}} }} \right)^{4}$ and
$N^{PXR}\sim ln\left( {\frac{{E_{p}} }{{mc^{2}}}} \right)$ at
$E_{p}
> > E_{tr} $. Note, such dependences of PXR intensity on energy
were calculated without taking into account of MS influence on
radiation characteristics.

The \textit{Bremsstrahlung} coherent length is directly
proportional to charged particle energy: $L_{Br} = \sqrt
{\frac{{4c}}{{\omega \overline {\theta _{s}^{2}} } }} $, where
$\overline {\theta _{S}^{2}}  = \frac{{1}}{{2}}\left(
{\frac{{E_{S}} }{{E_{P}} }} \right)^{2} \cdot \frac{{1}}{{L_{R}}
}$, \textit{E}$_{S} $= 21.2 MeV - scale energy, \textit{L}$_{R}$ -
radiation length. Therefore, by fixing the thickness of a crystal
target and changing the energy of charged particles (thus changing
$L_{Br} $), it is possible to calculate dependences of
spectral-angular density and angular distribution, and dependence
of spectral maximum width on the $L_{0} /L_{Br} $ ratio (similar
to shown in Fig.\ref{fig1} - Fig.\ref{fig3}). Certainly, the
nature of these dependences is completely different. A change of a
crystal target thickness results in the change of a trajectory
length where particle radiates coherently. In fact, this is its
cut-off because the PXR coherent length is equal to infinity.
Besides, on particle energy decrease the PXR generation is
realized at increasing values of $\left| {\alpha _{B}} \right|$
(detuning parameter from the exact value of the Bragg angle) that
in turn results in decrease of radiation intensity.

\begin{figure}
\epsfxsize = 8 cm \centerline{\epsfbox{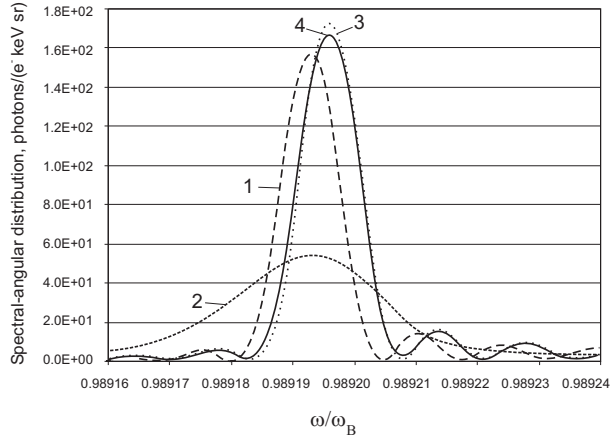}} \caption{PXR
spectral-angular distributions plotted for various energies of
electron beam: 1, 2 - 900 MeV; 3, 4-9 GeV; curves 1, 3 -
calculation without MS account; 2, 4 - calculation with MS
account} \label{fig4}
\end{figure}

PXR spectral-angular distributions calculated for the same
geometry as Fig.\ref{fig1} and Fig.\ref{fig2} were plotted in
Fig.\ref{fig4}.
%
%
%

Crystal thickness was chosen equal to 0.01 cm and polar angle of
radiation was $\vartheta $=1.8  mrad.
It is evident from Fig.\ref{fig4}, which for 9 GeV electron beam
energy distributions calculated without (curve 3) and with MS
account (curve 4), practically coincides.
In this case $L_{Br} = 1.22 \times 10^{ - 1}$cm, i.e. $L_{0} < <
L_{Br} $\textbf{,} so a situation of weak MS is realized here.
Change of energy from 9 GeV down to 900 MeV yields in change of
\textit{Bremsstrahlung} coherent length to the order of magnitude,
$L_{Br} = 1.22 \times 10^{ - 2}$cm\textbf{.} In this case target
thickness along direction of charged particle movement $L_{0} $
becomes of one order with $L_{Br} $. Presence of MS results in
essential decrease of height and spreading of PXR peak (curve 2)
in comparison with the calculations, without taking MS into
account (curve 1).

The spectrum corresponding to 900 MeV appeared shifted down for
$\Delta \omega \sim  5 \times 10^{ - 6}\omega _{B} $ along
frequency axis relative to spectrum corresponding to 9 GeV. In the
same time, shift of curve 2 relative to curve 1 as a result of MS
influence for 900 MeV appeared to be just $10^{ - 7}\omega _{B} $.

Amplitudes of spectral-angular distributions without MS (curves 1
and 3) has decreased less than 1.5 times, while MS account
decrease it more than four times with energy decrease from 9 GeV
down to 900 MeV. Comparing spectral-angular distributions for 900
MeV energy, it is possible to see that MS account results in
decrease of height of spectral-angular distribution approximately
3.2 times and its spreading is $\sim  $2.5 times. Thus, MS
presence results in decrease of PXR angular intensity.

\begin{figure}
\epsfxsize = 8 cm \centerline{\epsfbox{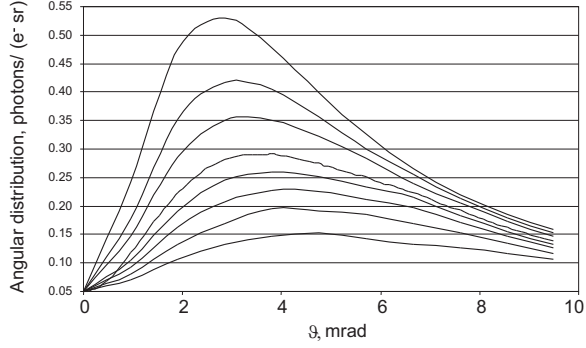}} \caption{PXR
angular density distributions at MS presence for various beam
energies. The lower curve corresponds to 300 MeV, further in upper
direction: 350, 400, 450, 500, 600, 700, and 900 MeV} \label{fig5}
\end{figure}

PXR angular distributions with MS account for the geometry
described above in a crystal with thickness of 0.13 cm are plotted
in Fig.\ref{fig5}
%
%
%
Detector resolution was considered equal to $\Delta \varpi /\varpi
_{B} = 10^{ - 3}$.
%
It is evident from the figure that energy decrease to three times
(from 900 down to 300 MeV) results in a decrease of maximum
intensity almost five times.
At the same time there is a shift of angular distribution maximum
to the bigger angles from 2.9  mrad up to 4.8   mrad and spreading
almost to one and half times more. Thus MS presence results in
noticeable decrease of angular intensity, angular distribution
spreading and to a shift of maximum in angular distribution
towards the bigger angles in comparison with calculations without
MS account. For example, for 300 MeV electron energy the maximum
in PXR angular distribution without MS account is achieved at 2.3
 mrad  angle.

Study of radiation yield energy dependence is important for
correct understanding of the PXR generation mechanism. This
dependence, as mentioned above, has a threshold nature and its
specific shape at experimental measurement in many respects is
determined by the angular size of the detector. As energy
decreases, the effective angle of radiation emission $\vartheta
_{ph} \approx \sqrt {\gamma ^{ - 2} - {\chi} '_{0} +
\frac{{\overline {\theta _{s}^{2}}  L_{eff}} }{{2}}} $ increases.
Value of \textit{L}$_{eff}$ depends on crystal thickness,
absorption length, Bragg angle, and radiation geometry \cite{8}.
%
%
%
\begin{figure}
\epsfxsize = 8 cm \centerline{\epsfbox{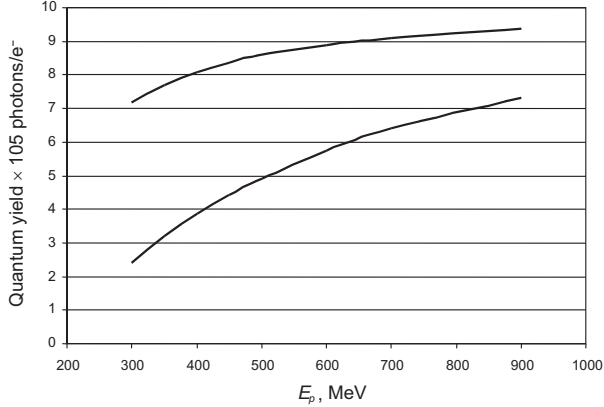}} \caption{PXR
quantum yield dependences on electron beam energy without MS
account (upper curve) and with MS account (lower curve)}
\label{fig6}
\end{figure}

Results of PXR quantum yield calculations for electron energy in a
range from 300 up to 900 MeV are plotted in Fig.\ref{fig6};
angular aperture of detector is 9.5$ \cdot $10$^{-3}$ rad. The
angular aperture of detector is $\theta _{D} > \vartheta _{ph} $
even for 300 MeV, that is why the detector collects practically
all generated radiation and we have behaviour of quantum yield
energy dependence that is ``natural'' for proper PXR.

Presence of MS results in a steep decrease of PXR quantum yield.
For example, for 700 MeV electron energy PXR quantum yield without
MS exceeds the result received with MS consideration by up to 1.5
times, for 300 MeV this value already amounts $\sim  $2.9 times.

\section{Conclusion}

\quad Summing up, multiple scattering of charged particles in a
crystal target changes considerably the characteristics of x-rays
originating due to a particle passage through a single crystal. MS
effect is bigger at low energies and/or thick targets and
demonstrates itself in x-rays distributions spreading,
monochromaticity, and quantum yield decrease. Nevertheless, PXR at
low energies can provide monochromaticity of 10$^{-3}$-10$^{-2}$
which is still applicable in medical imaging.

\begin{figure}
\epsfxsize = 8 cm \centerline{\epsfbox{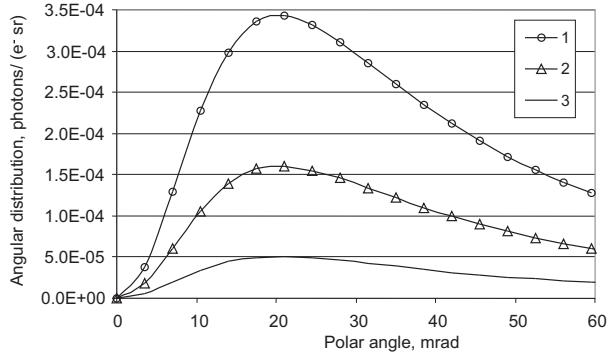}} \caption{Angular
density for 33 keV PXR emitted by 25  MeV electrons: 1 - for (111)
reflex, 2 - for (220) reflex, and 3 - for (400) reflex}
\label{fig7}
\end{figure}

Using the approach described above, let us estimate the
characteristics of PXR source, which can be produced on real
medical accelerators \cite{14}. As a rule, maximal energy of
electron beam in these accelerators is equal 20-28 MeV. We have
evaluated angular density of radiation for next conditions: E$_{e}
$= 25  MeV, 33 keV x-rays, symmetrical Laue case for (111), (220),
and (400). Silicon target of L=0.01 cm was chosen which is
somewhat bigger, but more real than optimal for these conditions
(about (1$\div$10)$L_{Br}$, i.e. $\sim 3 \div 30$ microns) target
thickness. Angles between electron velocity direction and
direction to diffraction reflex are $\sim  $6.9, 11.2, and 15.9
degrees, respectively. Angular densities for discussed cases are
plotted in Fig.\ref{fig7}.
%
%

Integration on angular density gives us total amount of x-ray
quanta on $20 \times 20$ cm area located at 1.5 m distance from
the target. Appropriate amounts appeared equal to $\sim 3 \cdot
10^{-6}$, $\sim 5 \cdot 10^{-7}$, and $\sim  1 \cdot 10^{-7}$
photons/e$^{-}$ for (111), (220), and (400) reflexes,
respectively. It is evident, that despite a decrease of radiation
quantum yield at low energies, the integral amount of quanta do
not decrease very drastically and the number of x-ray quanta
needful for quality image can still be achieved at 0.1-0.2 A beam
current.

However, to provide some margin of safety for PXR application for a medical
x-ray source, it is necessary to find opportunities to increase its
radiation yield. Let us consider some variants of radiation yield increase.
One of the ways to increase x-rays spectral-angular density may be
the generation of radiation in a multi-wave mode, in other words,
when PXR emission goes simultaneously on several systems of the
crystallographic planes. In \cite{15,16} measurements of angular
distributions of PXR generated in a GaAs crystal by relativistic
electrons with energies 500 and 900 MeV were reported. In these
experiments the anomalies (i.e. significant intensity increase in
narrow angular range), which cannot be explained by the PXR
two-wave theory \cite{10}, were observed. The analysis of geometry
has shown that conditions of multi-wave diffraction were realized
for emitted photons. Thorough theoretical description for
experiments \cite{15,16} was performed in \cite{17}. In \cite{18}
demonstration of multi-wave effects in low-energy range was
numerically analysed for PXR generated by 7 MeV electrons in
conditions of eight-wave diffraction. Calculations have shown
that, despite of strong MS of electrons at low energies,
multi-wave effects may appear in PXR angular distribution and also
result in the formation of a strong narrow peak in the centre of
two-wave angular distribution.

Next a rather promising way for radiation spectral-angular density
increase is the application of stratified (multi-layered) crystal
targets \cite{19}. PXR intensity increases up to 7-8 times in
10-layer target made of relatively thin silicon crystal foils in
comparison with monolithic target of equivalent thickness have
been reported in the paper. At last, targets made of mosaic
crystals should be mentioned (for example, \cite{20}). They also
can provide an increase in the intensity of x-rays.

Research of the behaviour of sophisticated targets at low energies,
preferably at conditions that can provide existing medical accelerators, and
determination of their optimal characteristics in this range, may lead to a
practical design for a real medical monochromatic x-ray source. At the end,
our results were obtained for Laue geometry, but for Bragg geometry MS
influence can be somewhat weaker. Bragg geometry of PXR radiation can also
provide more intensive x-ray yield. It is intended to be a subject of our
further research.

\section*{Acknowledgements}

Authors are very grateful to Dr. O. Missevich for fruitful discussions on
x-ray imaging issues.


\begin{thebibliography}{999999}

\bibitem{1} M.L. Ter-Mikaelian, High Energy Electromagnetic Processes in Condensed
Media, New York: Wiley, 1972.

\bibitem{2} V.G. Baryshevsky and I.D. Feranchuk, Zh. Eksp. Teor. Fiz. 61 (1972) 944
(Sov. Phys. JETP 34 (1972) 502).

\bibitem{3} G.M. Garybyan and C. Yang, Zh. Eksp. Teor. Fiz. 61 (1972) 430 (Sov. Phys.
JETP 34 (1972) 495).

\bibitem{4} Yu.N. Adishchev, V.G. Baryshevsky, S.A. Vorobiev, V.A. Danilov, S.D. Pak,
A.P. Potylitsyn, P.F. Safronov, and I.D. Feranchuk, Pis'ma Zh.
Eksp. Teor. Fiz. 41 (1985) 295 (Sov. Phys. JETP 41 (1985) 361).

\bibitem{5} V. Ingal and E. Beliaevskaya, \underline
{http://www.xraysite.com/knowbase/phaseradiology.html}

\bibitem{6} The Physics of Medical Imaging / S. Webb (Ed.), Bristol: Hilger, 1978.

\bibitem{7} R.B. Fiorito, D.W. Rule, M.A. Pierstrup et al., Nucl. Instrum. Meth. B 79
(1993) 758.

\bibitem{8} V.P. Afanasenko, V.G. Baryshevsky, A.S. Lobko, V.V. Panov, and R.F.
Zuevsky, Nucl. Instrum. Meth. A334 (1993) 631.

\bibitem{9} I.D. Feranchuk, A. Ulyanenkov, J. Harada, and J.C.H. Spence, Phys. Rev. E
62, \#3 (2000) 4225.

\bibitem{10} I.D. Feranchuk and A.V. Ivashin, J. de Physique (Paris) 46 (1985) 1981.

\bibitem{11} O. Lugovskaya, Characteristics of parametric x-rays in conditions of
dynamical diffraction and multiple scattering, Ph.D. Thesis, NAS
Institute for Physics, Minsk, Belarus, 2003.

\bibitem{12} V.G. Baryshevsky, Channelling, Radiation and Reactions in Crystals at
High Energies, Minsk: Belarussian University Publ., 1982 (in
Russian).

\bibitem{13} V.G. Baryshevsky, A.O. Grubich, and Le Tien Hai, Sov. Phys. JETP 94
(1988) 51

\bibitem{14} E.A. Abramian, Industrial Electron Accelerators, Moscow:
Energoatomizdat, 1986 (in Russian).

\bibitem{15} V.P. Afanasenko, V.G. Baryshevsky, O.T. Gradovsky et al., Phys. Lett.
141A (1989) 311.

\bibitem{16} V.P. Afanasenko, V.G. Baryshevsky, S.V. Gatsikha et al., Sov. JETP.
Lett. 15 (1990) 242.

\bibitem{17} S.A. Stepanov, A.Ya. Silenko, A.P. Ulyanenkov, and I.Ya. Dubovskaya,
Nucl. Instrum. Meth. B 117 (1996) 55.

\bibitem{18} I.Ya. Dubovskaya, In: Basic and Applied Physical Studies (1986-2001) /
V. Baryshevsky (Ed.), Minsk: Belarussian University Publ., 2001
(in Russian).

\bibitem{19} Y. Takashima, K. Aramitsu, I. Endo et al., Nucl. Instrum. Meth. B 145
(1998) 25.

\bibitem{20} R.B. Fiorito, D.W. Rule, X.K. Maruyama et al., Phys. Rev. Lett. 71, \#5
(1993) 704.
\end{thebibliography}
\end{document}